# Using SMART for Customized Monitoring of Windows Services

Gregory A. Pluta     Larry Brumbaugh     William Yurcik

National Center for Supercomputing Applications (NCSA)
University of Illinois at Urbana-Champaign
*{gpluta,ljbrumb,byurcik}@ncsa.uiuc.edu*

**ABSTRACT**

We focus on examining and working with an important category of computer software called *Services*, which are provided as a part of newer Microsoft Windows operating systems. A typical Windows user transparently utilizes many of these services but is frequently unaware of their existence. Since some services have the potential to create significant problems when they are executing, it is important for a system administrator to identify which services are running on the network, the types of processing done by each service, and any interrelationships among the various services. This information can then be used to improve the overall integrity of both the individual computer where a questionable service is running and in aggregate an entire network of computers.

NCSA has developed an application called SMART (Services Monitoring And Reporting Tool) that can be used to identify and display all services currently running in the network. A commercial program called Hyena remotely monitors the services on all computers attached to the network and exports this information to SMART. SMART produces various outputs that the system administrator can analyze and then determine appropriate actions to take. In particular, SMART provides a color coordinated user interface to quickly identify and classify both potentially hazardous services and also unknown services.

## 1. Introduction

Applications, processes, programs, tasks, threads and services are included among the terms commonly used to identify various types of software running on a computer system. The six terms identify at least four different categories of objects – services are a distinct type from the other five terms. However, since every executing service is associated with a process, these two terms are related. Furthermore, processes have a relationship with tasks and threads. Consequently, relationships exist between services and processes, tasks, threads, etc. Section 2 gives an overview of the role of services provided by current Windows operating systems. Section 3 describes various utility programs that can be used to examine and modify the services running on a computer. The utilities include *msconfig*, *services.msc*, *tasklist*, *net* and *sc*. Examples of the actual output produced by each of the utilities is shown along with output from Windows Task Manager. Section 4 describes a commercial program called Hyena that can be used to retrieve information about the services running on all the computers in a network. Hyena serves as the interface between the networked computers and the SMART program. Section 5 describes and illustrates the SMART application, showing various outputs produced. A combination of color and size is used to help quantify the results. Section 6 concludes and identifies future enhancements to SMART that are primarily concerned with further improving the quality of the output displayed.

## 2. Introduction to Windows Services

*Services* are programs that are loaded and begin running when the computer is initially booted, even if nobody logs onto the computer. This is different from allowing a user to start a program by clicking its icon on the desktop or launching it from the *Startup Folder* under *All Programs*. It is also different from identifying programs that are to start automatically after a user logs on. For example, suppose a server must be started and available in order for clients to contact it. A service could be used to do this, although clearly there are also other ways it can be accomplished without using services. A Windows service is comparable to a UNIX daemon program, in that both run in the background and process requests from the network and from other programs. Services are not processes, although the two share some important interrelationships that are later examined in Section 3.3. Services are often very loosely defined as programs that run in the background. However, not all programs running in the background are services. For example, anti-virus programs are not services. Rather, a program must be explicitly categorized as a service.

A wide range of services are automatically included by Microsoft as part of Windows operating systems, such as Windows 2000 and Windows XP. [Bott] contains a detailed listing of the Microsoft services which includes a basic description and security recommendations for each one. An even more detailed listing is available on-line at [ServicesGuide]. In reality, if services were not available, a noticeable amount of the standard work done on a computer would not be possible. Services are also provided by non-Microsoft vendors. The examples in Section 3 were run on a computer where Tivoli and McAfee services were running as can be confirmed by the listing in Figure 1.

In addition, users can create their own services. Suppose a user wants a simple program that monitors changes to a file. If the program is made a service, it will begin executing prior to the first file change. Initially user services were written in C++ in order to guarantee a secure and robust application that scales well. With the advent of .NET, Visual Basic or even VBScript can now be used to create Windows Services that formerly required a combination of C++ and MFC programming expertise. [Patterson]. An example is given in [Thews] that illustrates how the use of services can positively impact the quality of user application programming. Thews describes the integration of services into .NET applications. Very basic application services including database connection pooling, event logging, auditing, etc. can be developed and deployed as a Windows Service that can then be utilized by any .NET applications without repeating the same code. This results in smaller programs. Hence, common functionality can be supplied to any application that needs a specific function performed. This works especially well for those services that are requested by applications such as the three basic services mentioned above.

**2.1 Examples of the Three Service Startup Types**

Each service has a Startup type of automatic, manual or disabled. A service in the automatic state begins to run when the computer is booted. Automatic services perform a function that is routinely required during processing. There are a few services that will also automatically stop when no longer needed. If a service will not be needed, it should be placed in a manual or disabled state. A service in the manual state can be started by Windows when needed. In fact, most services in manual mode will start up automatically when they are needed. However, disabled services cannot be run unless their startup status is changed. Disabled services usually have the potential to produce some negative effects that preclude executing them. Manual services fall somewhere in between automatic and disabled. In general required services should be automatic, never used or dangerous services should be disabled and all others should be manual. Some examples of services that are usually automatic include DNS Client, DHCP Client, Error Reporting, Event Log, Help, Print Spooler and Protected Storage.

Some examples of services that are usually disabled include *ClipBook* (an almost never used feature), *Alerter* (a feature that can be abused) and *Telnet* (a true security risk). The disabled setting stop a service from starting, even when it is needed. Errors in the *Event Viewer* will identify trying to start services in the disabled state. Some services, while disabled, will constantly issue complaining diagnostic messages. However, this action can be avoided if they are placed in a manual state. The service descriptions in [Bott] and [ServicesGuide] offer advice on which services should be in manual and which in disabled.

**3. Utilities that Examine and Modify the Services Running on a Computer**

Numerous methods are available for displaying information about services that exist on a Windows computer. Several are described in this section, but no all-inclusive list is given. Some are available only with Windows XP and others only with Windows 2000. There is considerable overlap of processing capabilities among the utilities. Some of the utilities are command line driven, while others run in their own window. The utilities that run in their own window can be started either from the command line or via mouse selection. To start a program from the command line, type the name of the program in a Run box that can be reached from the *Start* menu (*Start -> Run ->* program name). Table 1 contains a summary of the utilities discussed.

**Table 1.** Utilities for Managing Windows Services

| Utility Program | Runs in Window or at Command Line | Important Processing Details and Scope of Processing Operations |
|---|---|---|
| *msconfig* | Window | processes several important system components |
| *services.msc* or *compmgmt.msc* | Window | strictly a services utility<br>provides the most information and control |
| *tasklist* or *tlist* (Win2000) | Command Line | ties service and process information together<br>a multi-purpose utility |
| *net* | Command Line | start/stop/pause/etc services<br>a multi-purpose utility |
| *sc* | Command Line | create/delete/start/stop/etc services<br>a multi-purpose utility |

### 3.1 The MS System Configuration Utility *(msconfig)*

This utility can be run by typing msconfig in a Run box. In addition to displaying services information, msconfig also generates information on important system file data (*SYSTEM.INI*, *WIN.INI*, and *BOOT.INI*), programs to start after the user logs-on and the type of startup to perform. Select the Services tab to generate a listing of all services found on the computer. For each service the name of its Manufacturer (most are Microsoft Corporation), its Status (Running or Stopped) and its classification as Essential (Yes) or Non-Essential (blank) is displayed. A limited amount of modifications can be made to the Services listing. By clicking on one of the four column headings, the listing can be sorted using the values in that column. A second click reverses the order of the sort. For example, all of the Microsoft services can be listed together. Three additional possibilities are to display only the non-Microsoft services, to *Disable All* services and to *Enable All* services. Furthermore, a checkbox in front of each service can be used to enable/disable it. With this utility it is difficult to edit the information associated with a specific service. Figure 1 shows output produced by *msconfig* after sorting by *Service* name. (This will be changed to sort by manufacturer in reverse order.)

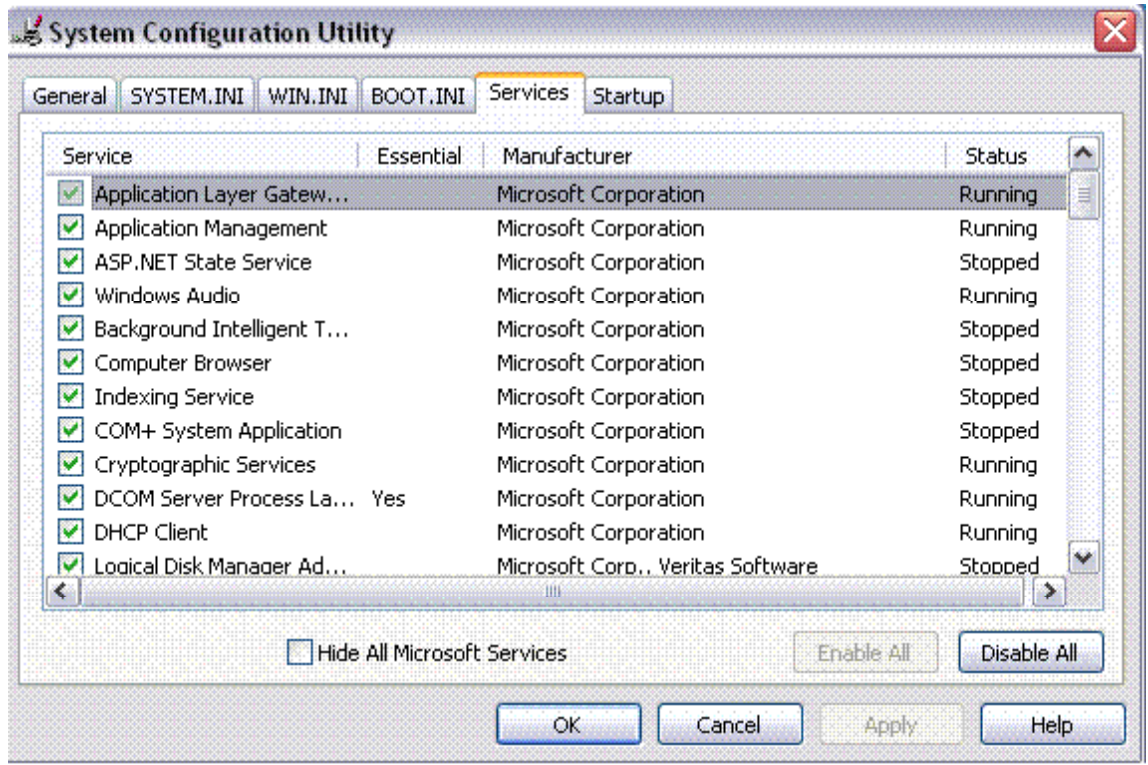

**Figure 1.** Output from the *msconfig* utility

**3.2 The Services Utility (services.msc)**

This utility provides more detailed services information than *msconfig*. Typing *services.msc* in a Run box displays the Services window. Alternatively, from the *Start Menu -> Control Panel -> Administrative Tools -> Services*. In addition, either right-clicking *My Computer* and then selecting *Manage* or typing *compmgmt.msc* at a command prompt will also produce the same display, called the *Services Console*. The utility provides a listing that for each service includes a moderately detailed *Description*, *Status* (Started or not-Started, which is blank), *Startup Type* (Manual, Automatic or Disabled) and *Log On As* (Local System, Local Service and Network Service). The *Log On As* value can usually be ignored. Figure 2 shows output produced by *services.msc*. There are two ways to make changes to an entry in the Services table. The simplest method is to double click on any entry in the line associated with a service. This displays the menu for the service. *Service Status* and *Startup Type* can be modified and parameters can be specified that are passed to the associated program that executes when the service is started. The path to the executable for the service is also displayed in the menu and it can be changed. The other way to modify services is to right click on a table entry and then select *Properties*. The same service menu is displayed as with the double click approach. The menu actually contains four tabs: (1) *General*, (2) *Log-on*, (3) *Recovery* and (4) *Dependencies*. Everything discussed above is under the *General* tab. *Recovery* identifies actions taken when a service fails. *Dependencies* identifies system components dependent on this service and also system components it is dependent upon. *Log-on* allows password protection.

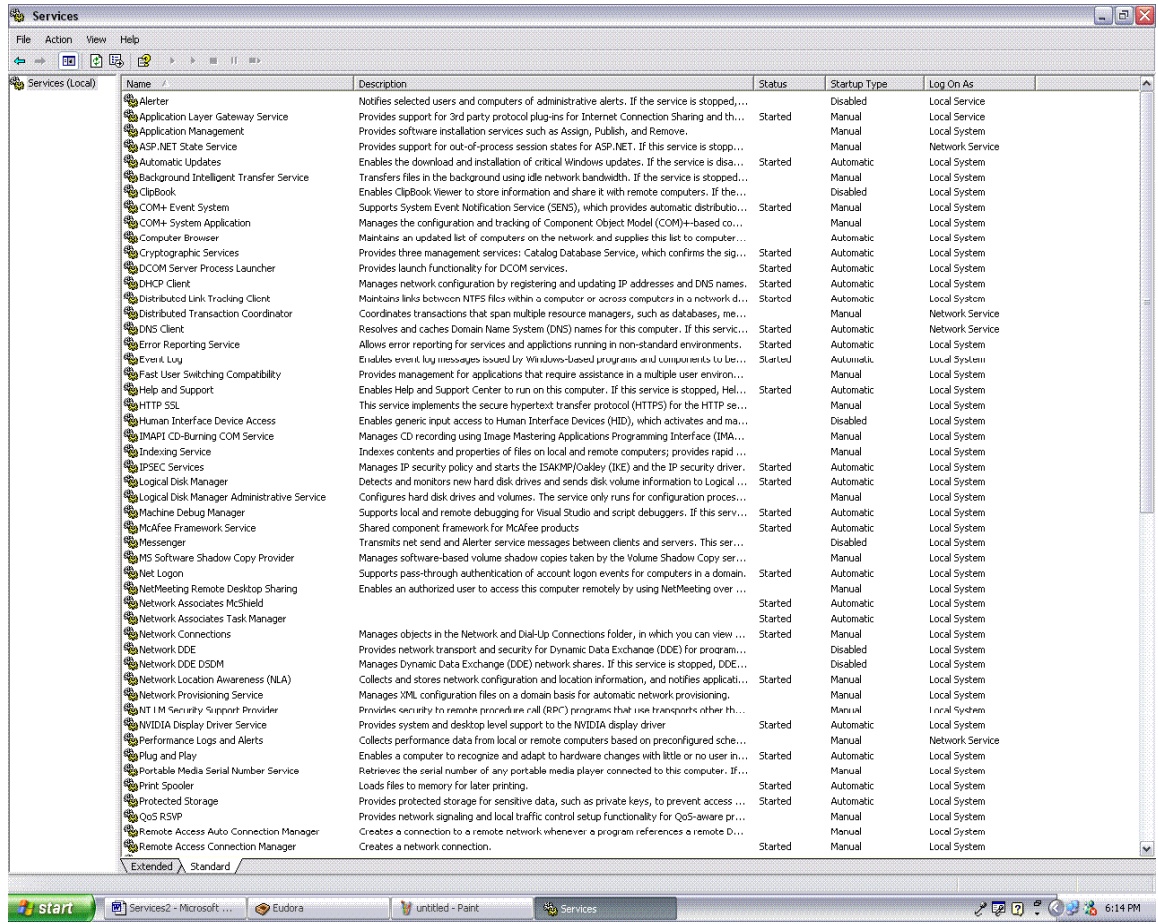

**Figure 2.** Output from services.msc utility

## 3.3 The *tasklist* Utility and the Relationship between Services and Processes

To generate a list of all processes running on a Windows computer use Ctrl-Alt-Delete to run *Task Manager*. Select the *Processes* tab and the currently executing processes are displayed. Most have an .exe extension. Figure 4 shows the output produced by the *Task Manager* under the *Processes* tab. The names of the processes have no correlation with the *Services* table entry names and there appears to be no apparent relationship between services and processes. However, as mentioned above the *services.msc* utility provides the name of the program that is run to implement a service. Using these names all of the automatic services have their associated program listed among the processes. If a manual, stopped service is started, a new process is created. Several services may use the same executable. In this case only one process entry ordinarily is listed. An exception is svchost.exe which is listed multiple times among the processes, all with different process IDs (PID). A considerable number of services have *svchost.exe* as their associated executable. For a complete listing of all this information showing the executables for all the services see [ServicesGuide].

However, there is a much easier way to determine the correlation between services and processes. The tasklist utility can be used to identify the association of services with the processes listed by task manager. Every service must be listed as associated with/part of a process. To produce such a listing type *tasklist /svc* at a command prompt. Figure 3 shows output produced by *tasklist /svc*. Here the *svc* flag is used to restrict the command to services.

The *svchost.exe* process requires some additional remarks. It is shown running multiple times concurrently by task manager and some of the occurrences have multiple services associated with them. *Svchost.exe* is a generic process name for services that run from dynamic-link libraries (DLLs). Multiple *svchost.exe* processes can run at the same time. Each *svchost.exe* session can contain multiple services. Different *svchost.exe* programs are required because some services require different information to be passed to them when they start. The grouping of services allows better control and simpler debugging.

```
Image Name                     PID Services
========================= ======= =======================================
System Idle Process             0 N/A
System                          4 N/A
smss.exe                      580 N/A
csrss.exe                     660 N/A
winlogon.exe                  684 N/A
services.exe                  728 Eventlog, PlugPlay
lsass.exe                     740 Netlogon, PolicyAgent, ProtectedStorage,
                                  SamSs
svchost.exe                   912 DcomLaunch
svchost.exe                   988 RpcSs
svchost.exe                  1080 AudioSrv, CryptSvc, Dhcp, dmserver, ERSvc,
                                  EventSystem, helpsvc, lanmanserver,
                                  lanmanworkstation, Netman, Nla, RasMan,
                                  Schedule, seclogon, SENS, SharedAccess,
                                  ShellHWDetection, srservice, TapiSrv,
                                  Themes, TrkWks, W32Time, winmgmt, wuauserv,
                                  WZCSVC
svchost.exe                  1128 Dnscache
svchost.exe                  1300 LmHosts, RemoteRegistry, SSDPSRV, WebClient
spoolsv.exe                  1448 Spooler
FrameworkService.exe         1616 McAfeeFramework
Mcshield.exe                 1664 McShield
VsTskMgr.exe                 1692 McTaskManager
naPrdMgr.exe                 1736 N/A
mdm.exe                      1780 MDM
nvsvc32.exe                  1840 NVSvc
dsmcad.exe                   1896 TSM Client Acceptor
dsmcsvc.exe                  1968 TSM scheduler
wdfmgr.exe                   2004 UMWdf
alg.exe                       484 ALG
explorer.exe                 1144 N/A
shstat.exe                    364 N/A
UpdaterUI.exe                 440 N/A
gcasServ.exe                  528 N/A
ctfmon.exe                    536 N/A
msmsgs.exe                    548 N/A
WZQKPICK.EXE                  624 N/A
gcasDtServ.exe                772 N/A
iexplore.exe                 1964 N/A
wuauclt.exe                  2964 N/A
WINWORD.EXE                  2508 N/A
MSOHELP.EXE                  3480 N/A
cmd.exe                      2824 N/A
sapisvr.exe                  2496 N/A
taskmgr.exe                  1280 N/A
wmiprvse.exe                 1584 N/A
cmd.exe                      3652 N/A
tasklist.exe                 3016 N/A
```

**Figure 3.** Sample Output from the *tasklist /svc* Utility

| Image Name | User Name | CPU | CPU Time | Mem Usage | Handles | Threads |
|---|---|---|---|---|---|---|
| taskmgr.exe | | 00 | 0:00:02 | 3,616 K | 60 | 3 |
| mspaint.exe | | 00 | 0:00:08 | 2,432 K | 170 | 6 |
| sapisvr.exe | | 00 | 0:00:00 | 6,212 K | 188 | 6 |
| svchost.exe | | 00 | 0:00:00 | 3,988 K | 131 | 6 |
| wuauclt.exe | | 00 | 0:00:00 | 4,876 K | 161 | 4 |
| gcasDtServ.exe | | 00 | 0:00:07 | 12,540 K | 340 | 5 |
| WZQKPICK.EXE | | 00 | 0:00:00 | 2,108 K | 27 | 1 |
| msmsgs.exe | | 00 | 0:00:00 | 3,728 K | 169 | 3 |
| ctfmon.exe | | 00 | 0:00:00 | 2,864 K | 66 | 1 |
| wdfmgr.exe | | 00 | 0:00:00 | 1,596 K | 65 | 4 |
| dsmcsvc.exe | | 00 | 0:00:00 | 7,604 K | 143 | 4 |
| dsmcad.exe | | 00 | 0:00:00 | 6,852 K | 128 | 5 |
| WINWORD.EXE | | 00 | 0:00:42 | 29,340 K | 406 | 9 |
| nvsvc32.exe | | 00 | 0:00:00 | 2,776 K | 88 | 3 |
| mdm.exe | | 00 | 0:00:00 | 2,524 K | 89 | 6 |
| naPrdMgr.exe | | 00 | 0:00:00 | 1,096 K | 104 | 5 |
| VsTskMgr.exe | | 00 | 0:00:00 | 348 K | 124 | 11 |
| Mcshield.exe | | 00 | 0:00:38 | 21,168 K | 193 | 18 |
| UpdaterUI.exe | | 00 | 0:00:00 | 380 K | 106 | 5 |
| FrameworkServic... | | 00 | 0:00:00 | 7,108 K | 288 | 11 |
| spoolsv.exe | | 00 | 0:00:00 | 4,520 K | 121 | 11 |
| svchost.exe | | 00 | 0:00:00 | 4,360 K | 215 | 16 |
| svchost.exe | | 00 | 0:00:00 | 3,080 K | 87 | 6 |
| Eudora.exe | | 00 | 0:01:44 | 7,892 K | 347 | 12 |
| svchost.exe | | 00 | 0:00:09 | 23,588 K | 1,521 | 68 |
| svchost.exe | | 00 | 0:00:00 | 4,108 K | 378 | 10 |
| svchost.exe | | 00 | 0:00:00 | 4,168 K | 142 | 7 |
| shstat.exe | | 00 | 0:00:00 | 500 K | 65 | 6 |
| gcasServ.exe | | 00 | 0:00:37 | 8,344 K | 179 | 5 |
| lsass.exe | | 00 | 0:00:00 | 880 K | 391 | 20 |
| services.exe | | 00 | 0:00:01 | 3,960 K | 291 | 16 |
| winlogon.exe | | 00 | 0:00:01 | 496 K | 569 | 20 |
| csrss.exe | | 00 | 0:00:08 | 1,492 K | 475 | 10 |
| explorer.exe | | 00 | 0:00:19 | 33,436 K | 915 | 12 |
| smss.exe | | 00 | 0:00:00 | 372 K | 21 | 3 |
| alg.exe | | 00 | 0:00:00 | 3,364 K | 112 | 6 |
| System | | 00 | 0:00:08 | 220 K | 303 | 66 |
| System Idle Process | SYSTEM | 99 | 5:36:40 | 16 K | 0 | 1 |

Processes: 38   CPU Usage: 0%   Commit Charge: 302M / 2461M

**Figure 4.** Sample Output from the *Task Manager*

### 3.4 The *net* Utility

Although the *tasklist* utility displays the services associated with a process, it cannot be used to modify information or properties of services. However, some of this type of processing can be done at the command line with the net command. At the command line, type: *net ACTION service-name*. Here *ACTION* is start, stop, pause or continue. Service-name is the actual name of a specific service. Some examples of output produced by the net utility are shown in Figure 5. In particular, typing *net start service-name* will cause service-name to start. Likewise, *net stop service-name* will cause the service to stop. Start and stop are used in the reverse order in Figure 5 with a service that is currently running. Pause and continue are not valid with some of the services. Finally, net is used for many functions unrelated to services.

```
C:\>net
The syntax of this command is:

NET [ ACCOUNTS | COMPUTER | CONFIG | CONTINUE | FILE | GROUP | HELP |
      HELPMSG | LOCALGROUP | NAME | PAUSE | PRINT | SEND | SESSION |
      SHARE | START | STATISTICS | STOP | TIME | USE | USER | VIEW ]

C:\>net stop dhcp
The DHCP Client service is stopping.
The DHCP Client service was stopped successfully.

C:\>net start dhcp
The DHCP Client service is starting.
The DHCP Client service was started successfully.

C:\>net pause dhcp
The requested pause or stop is not valid for this service.

More help is available by typing NET HELPMSG 2191.

C:\>_
```

**Figure 5**. Various Outputs Produced by the *net* Utility

### 3.5 The Service Controller Utility *(sc)*

*sc* is a command line program used for communicating with the NT Service Controller and services. It has some overlap with the net command, but it also provides the ability to create and delete services which none of the other utilities can do. Figure 6 shows output produced by the *sc* utility when no parameters are specified. As with most command line programs this produces a display of all of the available options. *sc* command includes options of create, create remotely, delete, start, stop, pause, continue, interrogate and query. Note that create and delete parameters access the registry. *sc* processes objects other than services. Running the *sc* query prompt at the bottom of the help screen supplies additional information about the utility.

```
C:\Documents and Settings\Larry>sc
DESCRIPTION:
        SC is a command line program used for communicating with the
        NT Service Controller and services.
USAGE:
        sc <server> [command] [service name] <option1> <option2>...

        The option <server> has the form "\\ServerName"
        Further help on commands can be obtained by typing: "sc [command]"
        Commands:
          query-----------Queries the status for a service, or
                          enumerates the status for types of services.
          queryex---------Queries the extended status for a service, or
                          enumerates the status for types of services.
          start-----------Starts a service.
          pause-----------Sends a PAUSE control request to a service.
          interrogate-----Sends an INTERROGATE control request to a service.
          continue--------Sends a CONTINUE control request to a service.
          stop------------Sends a STOP request to a service.
          config----------Changes the configuration of a service (persistant).
          description-----Changes the description of a service.
          failure---------Changes the actions taken by a service upon failure.
          qc--------------Queries the configuration information for a service.
          qdescription----Queries the description for a service.
          qfailure--------Queries the actions taken by a service upon failure.
          delete----------Deletes a service (from the registry).
          create----------Creates a service. (adds it to the registry).
          control---------Sends a control to a service.
          sdshow----------Displays a service's security descriptor.
          sdset-----------Sets a service's security descriptor.
          GetDisplayName--Gets the DisplayName for a service.
          GetKeyName------Gets the ServiceKeyName for a service.
          EnumDepend------Enumerates Service Dependencies.

        The following commands don't require a service name:
        sc <server> <command> <option>
          boot------------(ok | bad) Indicates whether the last boot should
                          be saved as the last-known-good boot configuration
          Lock------------Locks the Service Database
          QueryLock-------Queries the LockStatus for the SCManager Database
EXAMPLE:
        sc start MyService
Would you like to see help for the QUERY and QUERYEX commands? [ y | n ]:
```

**Figure 6.** Output Produced by the *sc* Utility

**4. The Hyena Software**

The Hyena program is available from AMT software. It performs a variety of functions, only one of which is relevant to this discussion. Hyena contains a standalone component called SystemTools Exporter Pro that can be used to create delimited text output files of many different types of system information including Active Directory information, computers, disk space, groups, printers, network configuration, registry, scheduled tasks, shares, user rights and security information, users, WMI, and services. Exporter Pro also supports exporting and consolidation of logon-related information (such as user last logon) from all domain controllers. The delimited text files created by Exporter Pro can be imported into any database or spreadsheet program. The SMART program (to be described) exports services information into an Access data base. Hence, it has knowledge of all services on all machines in the network.

Hyena allows all of the processing with services listed in Table 2 to be performed. This is the same types of processed performed with the MS *System Configuration Utility and*

*Administrative Tools -> Services*. However the operations can be performed remotely. Obviously, Hyena can gather information about the services on any computer.

**Table 2.** Types of Services Processing Performed Remotely by Hyena

> Fast viewing and sorting of all service properties on any computer
> Service control: start, stop, pause, continue, or restart a service
> View service dependencies
> Service installation on multiple computers
> Service removal
> Modify service properties
> Control services on multiple computers at the same time
> Change service startup information, including password, on multiple computers
> Remote viewing of device drivers
> Full support for Windows 2000 service features, including Recovery options

## 5. The SMART (Services Monitoring And Reporting Tool) Application

The SMART program is an ASP application and is written in VBScript. It processes one or two Access databases to perform its functions. When the SMART application runs it takes the services information that was exported from Hyena and uses it to produce an output listing of all services found on the network as shown in Figure 7.

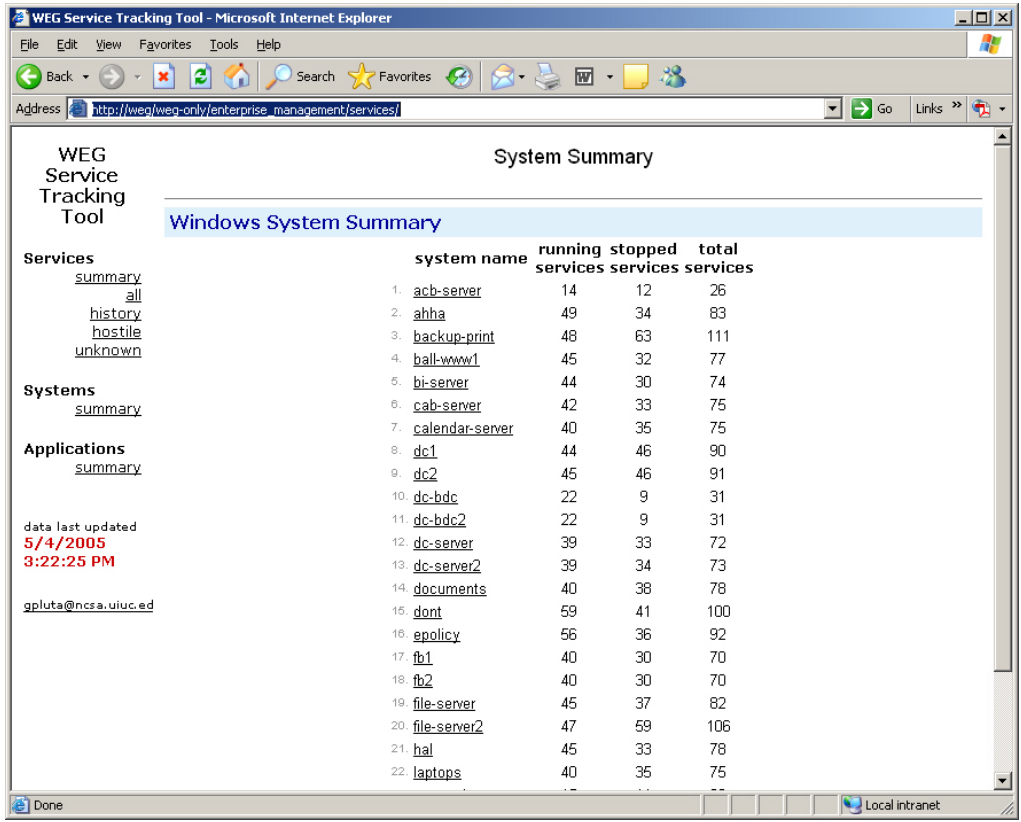

**Figure 7.** SMART Output: Windows Services Grouped by Individual Systems

The Services are then grouped into three categories and color coded as shown in Figure 8. Known hostile services (known to be used by hackers) are listed first in red and denote services that should not be running. Clicking on a particular service will show which systems are running the service. This is the highest priority information in the report. The system administrator will contact the persons running these services. Yellow entries are listed next and denote unknown services. An attempt will be made to try and identify the type of processing these services perform and add them to a database of all know services encountered on the network. Finally green entries denote known services that are categorized as permissible to run. This includes most of the default services provided by Microsoft. Currently the green entries are not displayed.

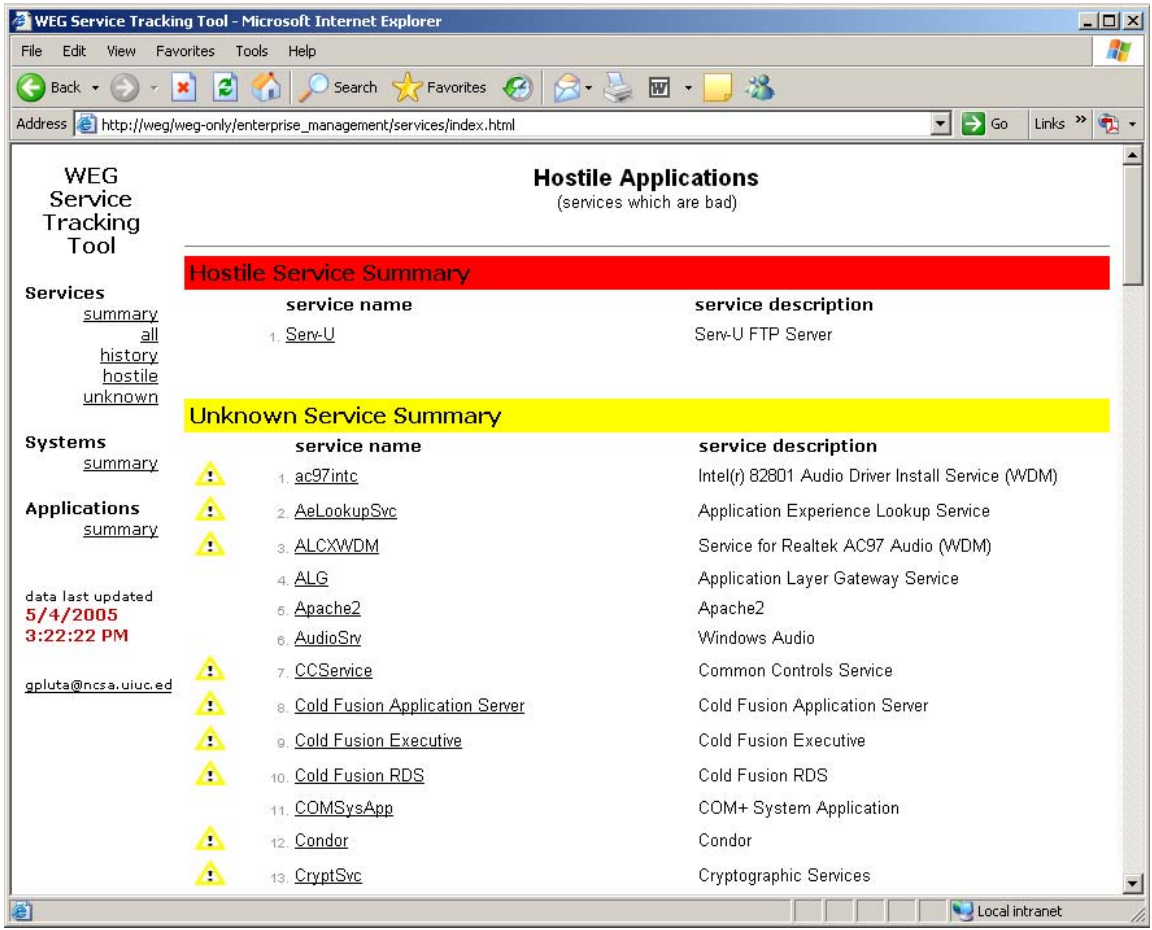

**Figure 8.** SMART Output: Unknown and Hostile Services

If a specific service entry is clicked, a report is generated that for each service identifies the number of computers on which it is running and the number on which it is stopped. These two numbers are also added together to produce a total. Table 9 shows SMART output for all the services aggregated over an entire network.

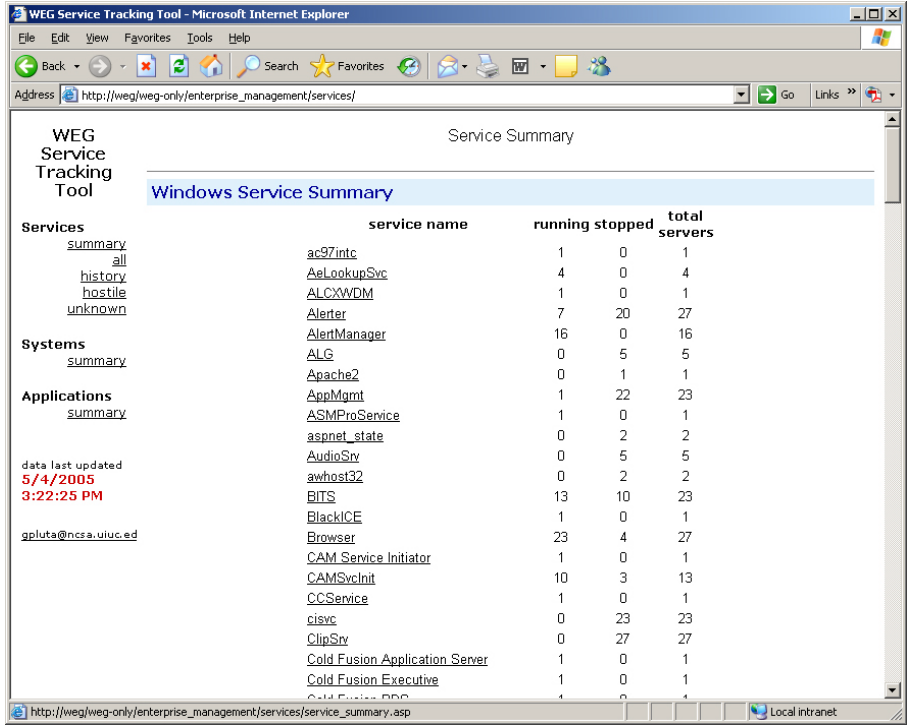

**Figure 9.** SMART Output: Windows Services Listing for An Entire Network

Figure 10 shows all services associated with a particular application. Also, for each application, a brief description of its function and a path to its executable are given.

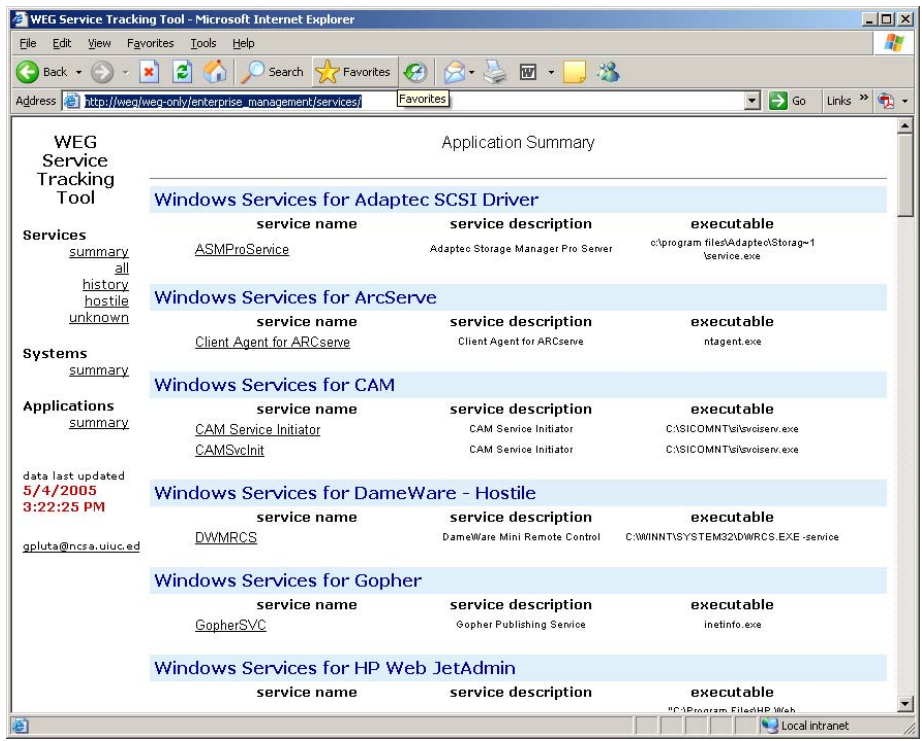

**Figure 10.** SMART Output: Applications View with all Associated Services

## 6. Conclusions and Future Enhancements

The SMART program provides the unique capability for tracking Windows services on a network of computers. This capability is novel and promises to have a positive impact on both network and security management since services are both functional communication channels as well as entry points for attackers. Being aware of the state of Windows services will allow system administrators to more quickly determine when operations are normal or abnormal requiring further investigation.

We are working on improvements to the SMART GUI display in order to better graphically represent multidimensional SMART output to the user. Using visualization techniques promises to make it easier for system administrators to more quickly browse and comprehend available information in order to make inferences based upon it.

## References

[Bott] Ed Bott and Carl Siechert, *Microsoft Windows Security Inside Out for Windows XP and Windows 2000,* Microsoft Press, 2003.

[Patterson] Brian Patterson, William Sempf, Richard Conway and Robin Dewson, *Visual Basic .NET Windows Services Handbook (1$^{st}$ edition)*, Wrox Press, 2002.

[ServicesGuide] *Services Guide for Windows XP.*
    <http://www.theeldergeek.com/services_guide.htm>

[Thews] Doug Thews, "Build Your First .Net Windows Service," *VisualStudio Magazine*, November 2004.